\documentclass[longauth]{aa}
\usepackage{natbib}
\bibpunct{(}{)}{;}{a}{}{,}
\usepackage{graphicx}
\usepackage{txfonts}

\begin{document}

\title{Amplified radio emission from cosmic ray air showers in thunderstorms}

\author{S.~Buitink\inst{1}
\and W.D.~Apel\inst{2}
\and T.~Asch\inst{9} 
\and F.~Badea\inst{2}
\and L.~B\"ahren\inst{3}
\and K.~Bekk\inst{2} 
\and A.~Bercuci\inst{4}
\and M.~Bertaina\inst{5} 
\and P.L.~Biermann\inst{6}
\and J.~Bl\"umer\inst{2,7}
\and H.~Bozdog\inst{2}
\and I.M.~Brancus\inst{4}
\and M.~Br\"uggemann\inst{8}
\and P.~Buchholz\inst{8}
\and H.~Butcher\inst{3}
\and A.~Chiavassa\inst{5}
\and F.~Cossavella\inst{2}
\and K.~Daumiller\inst{2} 
\and F.~Di~Pierro\inst{5}
\and P.~Doll\inst{2} 
\and R.~Engel\inst{2}
\and H.~Falcke\inst{1,2,6}
\and H.~Gemmeke\inst{9} 
\and P.L.~Ghia\inst{10}
\and R.~Glasstetter\inst{11} 
\and C.~Grupen\inst{8}
\and A.~Haungs\inst{2} 
\and D.~Heck\inst{2} 
\and J.R.~H\"orandel\inst{7} 
\and A.~Horneffer\inst{1}
\and T.~Huege\inst{2}
\and K.-H.~Kampert\inst{11}
\and Y.~Kolotaev\inst{8}
\and O.~Kr\"omer\inst{9}
\and J.~Kuijpers\inst{1}
\and S.~Lafebre\inst{1}
\and H.J.~Mathes\inst{2} 
\and H.J.~Mayer\inst{2} 
\and C.~Meurer\inst{2}
\and J.~Milke\inst{2} 
\and B.~Mitrica\inst{4}
\and C.~Morello\inst{10}
\and G.~Navarra\inst{5}
\and S.~Nehls\inst{2}
\and A.~Nigl\inst{1}
\and R.~Obenland\inst{2}
\and J.~Oehlschl\"ager\inst{2} 
\and S.~Ostapchenko\inst{2} 
\and S.~Over\inst{8}
\and M.~Petcu\inst{4} 
\and J.~Petrovic\inst{1}
\and T.~Pierog\inst{2} 
\and S.~Plewnia\inst{2}
\and H.~Rebel\inst{2} 
\and A.~Risse\inst{13} 
\and M.~Roth\inst{7} 
\and H.~Schieler\inst{2} 
\and O.~Sima\inst{4} 
\and K.~Singh\inst{1}
\and M.~St\"umpert\inst{7} 
\and G.~Toma\inst{4} 
\and G.C.~Trinchero\inst{10}
\and H.~Ulrich\inst{2}
\and J.~van~Buren\inst{2}
\and W.~Walkowiak\inst{8}
\and A.~Weindl\inst{2}
\and J.~Wochele\inst{2} 
\and J.~Zabierowski\inst{13}
\and J.A.~Zensus\inst{6}
\and D.~Zimmermann\inst{8}
}

\authorrunning{Buitink et al}
\titlerunning{Radio emission of air showers in thunderstorms}

\institute{Radboud University Nijmegen, Department of Astrophysics, IMAPP,
           P.O. Box 9010, 6500 GL Nijmegen, The~Netherlands\\
              \email{S.Buitink@astro.ru.nl}
             \and
            Institut\ f\"ur Kernphysik, Forschungszentrum Karlsruhe,
            76021~Karlsruhe, Germany\\
	    \and
	    ASTRON, 7990 AA Dwingeloo, The Netherlands \\
	    \and
	    Horia Hulubei National Institute of Physics and Nuclear Engineering (IFIN-HH),
            077125~Magurele-Bucharest, Romania\\
            \and
	     Dipartimento di Fisica Generale dell'Universit{\`a},
             10125 Torino, Italy\\
\and
Max-Planck-Institut f\"ur Radioastronomie,
53010 Bonn, Germany \\
\and
Institut f\"ur Experimentelle Kernphysik,
Universit\"at Karlsruhe, 76021 Karlsruhe, Germany\\
\and
Fachbereich Physik, Universit\"at Siegen, 57068 Siegen, 
Germany \\
\and
Inst. Prozessdatenverarbeitung und Elektronik, 
Forschungszentrum Karlsruhe, 76021~Karlsruhe, Germany \\
\and
Istituto di Fisica dello Spazio Interplanetario, INAF, 
10133 Torino, Italy \\
\and
Fachbereich Physik, Universit\"at Wuppertal, 42097
Wuppertal, Germany \\
\and
Radioastronomisches Institut der Universit\"at Bonn, 
53121 Bonn, Germany \\
\and
Soltan Institute for Nuclear Studies, 90950~Lodz, 
Poland\\
}
\date{Received; accepted} 
\offprints{S.Buitink@astro.ru.nl}
\abstract{
The detection of radio pulses from cosmic ray air showers is a potentially powerful new detection mechanism for studying 
spectrum and composition of ultra high energy cosmic rays that needs to be understood in greater detail.
The radiation consists in large part of geosynchrotron radiation. The intensity of this radiation depends, among other factors,
on the energy of the primary particle and the angle of the shower axis with respect to the 
geomagnetic field.}
{Since the radiation mechanism is based on particle acceleration, the atmospheric electric field can play an important role. Especially inside
thunderclouds large electric fields can be present. In this paper we examine the contribution of an electric field to the emission mechanism
theoretically and experimentally.}
{Two mechanisms of amplification of radio emission are considered: the
acceleration radiation of the shower particles and the radiation from the current that is produced by ionization electrons moving 
in the electric field. For both mechanisms analytical estimates are made of their effects on the radio pulse height.
We selected \textsc{lopes} data recorded during thunderstorms, periods of heavy cloudiness and periods of cloudless weather. We tested whether the 
correlations with geomagnetic angle and primary energy vary with atmospheric conditions.}
{We find that during thunderstorms the radio emission can be strongly enhanced. The present data suggests that the observed amplification is caused by acceleration of the shower electrons and
positrons. In the near future, extensions of \textsc{lopes} and the construction of \textsc{lofar} will help to identify the mechanism in more detail.
No amplified pulses were found during periods of cloudless sky or heavy cloudiness, suggesting that the electric field effect for radio air shower
measurements can be safely ignored
during non-thunderstorm conditions.}{}
\maketitle

\section{Introduction}                             \label{sec:introduction}

The first detection of radio pulses from extensive air showers (EAS) was in 1964 \citep{J65} and several emission mechanisms have 
been proposed to explain them. \citet{A62} calculated the Cherenkov
radiation resulting from the negative charge excess in extensive air showers. \citet{KL66} considered two more mechanisms, both driven by the geomagnetic field. Firstly, the 
geomagnetic field separates the negative and positive charges. The electric dipole created in this way will emit Cherenkov-like radiation in the atmosphere. Secondly, the
transverse current that is generated by the charge separation produces a radiation field which is strongly beamed forwards in the direction of the EAS. They predicted
the latter mechanism to be dominant. 

The atmospheric electric field also contributes to the total radio emission. \citet{Ch67} calculated the effect of charge separation due to electric fields and concluded that
this effect can contribute significantly when a large electric field is present in the atmosphere. Furthermore, \citet{W57} and \citet{Ch68} suggested 
that ionization electrons, left behind by the EAS, emit radiation when accelerated in a background electric field. Inside thunderclouds the electric field can be large
enough to accelerate the
ionization electrons up to energies high enough to produce ongoing ionization (typically 100 kVm$^{-1}$).
This effect is called runaway breakdown and a calculation of the associated radio pulse is done by \citet{G02}.

In the 1970's several groups did experiments with shower arrays and radio antennas. One of their aims was to infer the correct emission mechanism from polarization measurements.
Although most experiments were in favour of the transverse current mechanism, a large contribution of other mechanisms could not be excluded. 
Large spreads in radio
intensity, inability to filter out radio interference, and the significant but unknown effects of atmospheric conditions led to the abandonment of these experiment \citep[see][]{B77}.
Excessively large radio pulses of EAS during
thunderstorms were found experimentally by \citet{M74}.

Recently, \citet{G04} reported the discovery of radio pulses of duration $\sim 0.5$ $\mu$s associated with EAS during thunderstorms at the Tien Shan Scientific Station in Kazakhstan.
These were detected with antennas sensitive to frequencies between 0.1 and 30 MHz. The \textsc{forte} satellite has detected strong intracloud radio pulses in the 26--48~MHz
band, which are correlated with discharge processes inside the thundercloud \citep{J03}. 

The development of \textsc{lofar} revived the interest in radio detection of EAS. \citet{FG03} describe the emission in terms of coherent geosynchrotron
emission. Although no explicit comparisons are made, they expect this mechanism to be largely equivalent to the transverse current mechanism, since it also results from charge
separation and is beamed forward. Detailed simulations of the geosynchrotron emission from EAS are presented in \citet{HF03} and \citet{HF05}. 

The \textsc{lopes} experiment was set up at the \textsc{kascade} array site in Karlsruhe, consisting of 10 and later 30 radio antennas sensitive
to radiation in the band 40--80 MHz. The presence of the \textsc{kascade} particle detectors on the site, new techniques and
modern hardware allow higher resolution and a higher detection rate than the old experiments. In \citet{F05} it was shown that 
the measured antenna electric field is strongly correlated with the
muon number and the angle
of the shower axis with the geomagnetic field. The first correlation is a strong indication that the radio emission from EAS is coherent, while the latter correlation 
proves that the dominant emission mechanism is of geomagnetic nature. 

In this paper we investigate theoretically and experimentally the effect of atmospheric conditions on the radio emission. We compare sets of events that were recorded by \textsc{lopes} under various weather
conditions: fair weather, large nimbostratus clouds, and thunderstorms. Under violent weather conditions, the effect of geoelectric mechanisms increases. When this increase is large
enough it may dominate over the geomagnetic emission mechanism. Understanding the effect of atmospheric conditions on the emitted radiation
is crucial for a correct determination of the primary energy from observed radio pulses.

\section{Electric fields inside clouds}                  \label{sec:clouds}
We present some general characteristics of thunderstorms based on the information provided by \citet{MR98}.
In fair weather, i.e. atmospheric conditions in which electrified clouds are absent, there is a downward electric field present with a field strength of $\sim 100$\,Vm$^{-1}$ at ground
level. The field strength decreases rapidly with altitude and has values below 10~Vm$^{-1}$ at altitudes of a few hunderd meter and higher. The associated fair weather current charges up cloud boundaries, because clouds have lower conductivity than the free
atmosphere. Other effects, such as ion capture and collisions between polarized cloud particles, also contribute to the charging up of clouds. Clouds can typically gain field 
strengths of a few hundred Vm$^{-1}$. Nimbostratus clouds, which have a typical
thickness of more than 2000\,m can have fields of the order of 10\,kVm$^{-1}$. The largest electric fields are found inside thunderstorms, where locally field strengths can reach
values up to 100\,kVm$^{-1}$.
In most clouds this field is directed vertically (either upwards or downwards, depending on the type of cloud), but thunderclouds contain complex charge distributions and 
can have local fields in any direction. Thunderclouds can have a vertical extent of $\sim$ 10 km.
The electric field at ground level is strongly affected by the electric processes inside thunderclouds. Although it can not be used to estimate the field strength inside the cloud, a
change in the (polarity of) the ground field is a strong indication that large electric fields are present overhead. In the context of EAS radio emission, ground level
electric field mills can be used as a warning system for violent electric phenomena in the atmosphere. 

\section{Electric field influence on radiation}       \label{sec:influence}

The atmospheric electric field acts on the radio emission from EAS in various ways. We distinguish two generations of electrons: the relativistic electrons from pair
creation in the EAS (called shower electrons from here) and the non-relativistic electrons resulting from the ionization of air molecules by the EAS particles (called ionization
electrons from here). The shower electrons are created together with an equal amount of positrons. In this section, we will use Gaussian units. 
\begin{enumerate}
\item The electric force accelerates the shower electrons and positrons, producing radiation in more or less the same way as the magnetic field does. This effect 
is described in
Sec. \ref{subsec:showerelec}.
\item The ionization electrons are accelerated in the electric field. A radio pulse will be emitted from the short current that is produced in this
way. This effect is described in Sec. \ref{subsec:ionizedelec}.
\item As the electrons and positrons move through the electric field they can gain or lose energy, which has an influence on the electromagnetic cascade. When the electric field
points upwards and the shower is vertical the electrons gain as much energy as the positrons lose. However, when a charge excess of $\epsilon$ exists
a shower with primary energy $E_{0}=10^{16}$ eV and electron number $N\sim 10^{7}$ moving through a
thundercloud sustaining a 1 kV/cm vertical electric field over 2 km, the energy gain of the whole shower will be:
\begin{equation}
\Delta E= \epsilon N q |\vec{E}| \Delta x \approx 2\cdot10^{14} \mbox{eV}
\end{equation}
where $|\vec{E}|$ is the electric field strength, and we have used a typical value of $\epsilon=0.08$. This corresponds to $\sim$2\% of the 
primary energy. This value is only a very rough approximation.
 \item The electric field acts on all charged particles in the EAS and can thus influence the longitudinal and lateral development of the EAS. 
This may affect the coherence and the shape and size of the radio footprint. 
\end{enumerate}
The last two effects can best be investigated by including electric field effects in a Monte Carlo code like \textsc{corsika} \citep{H98} and are not studied in this paper. 
The first two effects will be further explored below. Although both mechanisms can be responsible for a strong enhancement of radio emission, they can be distinguished by their
temporal and spectral radiation profiles.

\subsection{Acceleration of air shower electrons} \label{subsec:showerelec}
The radiation part of the electric field of a moving electric charge can be expressed as \citep{Jackson}:
\begin{equation}
\label{radfield}
\vec{E}(\vec{x},t)=\frac{e}{c}\left[ \frac{\vec{n}\times\left[ (\vec{n}-\vec{\beta})\times\vec{\dot{\beta}} \right] }{(1-\vec{\beta}\cdot\vec{n})^{3}R}\right]_{\mathrm{ret}}
\end{equation}
where $e$ is the unit charge, $\vec{R}$ is the distance to the observer, $\vec{\beta}=\vec{v}/c$ is the velocity of the charge and $\vec{n}$ is a unit vector pointing in the
direction of the observer (approximating the index of refraction for radio waves in the atmosphere to be unity). The associated vector potential is:
\begin{equation}
\label{AfromE}
\vec{A}(\vec{x},t)=\left(\frac{c}{4\pi}\right)^{1/2}[R\vec{E}]_{\mathrm{ret}}
\end{equation}

\begin{figure}[h]
\begin{center}
\centerline{\includegraphics*[width=0.5\columnwidth]{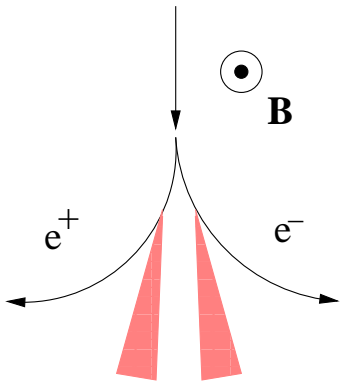}}
\caption{\label{geosynchr} The electrons and positrons travel along curved trajectories in the magnetic field, emitting synchrotron radiation.}
\end{center}
\end{figure}

As the electron and positron paths are curved in the geomagnetic field they emit synchrotron radiation as in Fig. \ref{geosynchr}. 
From Eqns. \ref{radfield} and \ref{AfromE} the vector potential in frequency domain can be calculated \citep{HF03}:
\begin{equation}
\label{vectorpot}
\vec{A}(\vec{R},\omega)=\frac{\omega e}{\sqrt{8 c}\pi}\mathrm{e}^{i(\omega\frac{R}{c}-\frac{\pi}{2})}\left[-\hat{e}_{\parallel}
A_{\parallel}(\omega)\pm\hat{e}_{\perp}A_{\perp}(\omega)\right]
\end{equation}
where the plus-sign corresponds to electrons and the minus-sign to positrons, $\omega$ is the angular frequency of the
radiation, $\hat{e}_{\parallel}$ the unit vector in the plane of curvature of the trajectories and $\hat{e}_{\perp}$ the unit vector perpendicular to that plane, both transverse
to $\vec{n}$. The components are given by:
\begin{equation}
A_{\parallel}=i\frac{2\rho}{\sqrt{3}c}\left(\frac{1}{\gamma^2}+\theta^{2}\right) K_{2/3}(\xi)
\end{equation}
\begin{equation}
A_{\perp}=\theta\frac{2\rho}{\sqrt{3}c}\left(\frac{1}{\gamma^2}+\theta^{2}\right)^{1/2} K_{1/3}(\xi)
\end{equation}
with
\begin{equation}
\xi=\frac{\omega \rho}{3c} \left(\frac{1}{\gamma^2}+\theta^{2}\right)^{3/2}
\end{equation}
where $K_{a}$ is the modified Bessel-function of order $a$, $\theta$ is the (small) angle between $\vec{n}$ and $\vec{\beta}$, and the radius of curvature $\rho$ is given by:
\begin{equation}
\rho=\frac{v\gamma m_{e} c}{e B \sin\alpha}
\label{radofcurv}
\end{equation}
where $v$ and $\gamma$ are the speed and Lorentz factor of the emitting particle, $B$ is the magnetic field strength and $\alpha$ is the angle of the trajectory with the magnetic
field direction (the pitch angle).
It can be seen from Eqn. \ref{vectorpot} that 
although the
radio emission components of the electron and positron perpendicular to the plane of curvature cancel out, the components in the plane of curvature add up. 

The radiated power of an accelerated charge can be expressed in the following form, known as the Li\'enard result \citep{Jackson}:
\begin{equation}
P=\frac{2}{3}\frac{e^2}{c}\gamma^{6}\left[ (\dot{\vec{\beta}})^{2} - (\vec{\beta}\times\dot{\vec{\beta}})^2\right]
\end{equation}
from which it can be seen that the power is related to the acceleration as $P_{\parallel} \propto \gamma^6 \dot{\vec{\beta}}_{\parallel}^2$ in case of linear acceleration and
$P_{\perp} \propto \gamma^4 \dot{\vec{\beta}}_{\perp}^2$ when the acceleration is perpendicular to the direction of motion. Furthermore, $\dot{\vec{\beta}}_{\parallel} \propto
\gamma^{-3} F_{\parallel}$ and $\dot{\vec{\beta}}_{\perp} \propto \gamma^{-1} F_{\perp}$, where $F_{\parallel}$ is a force in the direction of motion and $F_{\perp}$ is a force
perpendicular to the direction of motion.
Comparing the effect of these two forces one finds:
\begin{equation}
P_{\parallel}=\frac{2}{3} \frac{e^{2}}{m^{2}c^{3}} F_{\parallel}^{2}
\end{equation}
while:
\begin{equation}
P_{\perp}=\frac{2}{3} \frac{e^{2}}{m^{2}c^{3}}\gamma^{2} F_{\perp}^{2}
\end{equation}
which is a factor $\gamma^{2}$ greater.
The Lorentz force is always perpendicular to the particles' direction, but the electric force
can have any angle with respect to the trajectory.

\begin{figure}[h]
\begin{center}
\centerline{\includegraphics*[width=0.4\columnwidth]{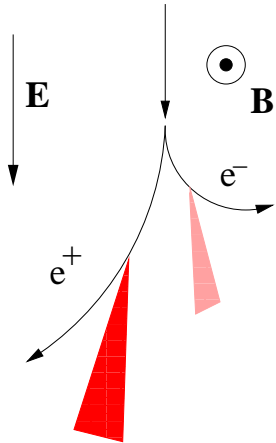}}
\caption{\label{trajectories} The electrons and positrons make curved trajectories in the magnetic field. Under influence of a downward
directed electric field, the positrons (electrons) are accelerated (decelerated). The asymmetry in the 
trajectories will be reflected in the radio emission.}
\end{center}
\end{figure}
\begin{figure}[h]

\begin{center}
\centerline{\includegraphics*[width=0.5\columnwidth]{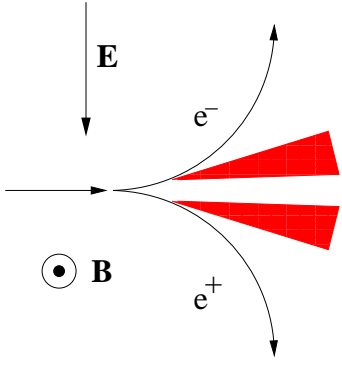}}
\caption{\label{trajectories2} Electron/positron pairs are created in a horizontal shower. The downward electric force works in the same direction as the Lorentz force for 
both species. The radio emission is amplified.}
\end{center}
\end{figure}
For a more detailed view we turn our attention to the vector potential. We will treat a general case in which a linear force and a
perpendicular force act on a charge and evaluate the vector potentials caused by these two forces. We define the unit vector $\hat{e}_1$ as
perpendicular to the particle's velocity $\vec{\beta}$ and lying in the orbital plane of the particle. From Eqn. \ref{vectorpot} we know that the emission
from an electron/positron pair is strongly polarized in this direction. We therefore evaluate the vector potentials of linear and
perpendicular acceleration along this unit vector for an observer in the orbital plane.
Since both the charge and $\dot{\vec{\beta}}$ in Eqn. \ref{radfield} have
different signs for electrons and positrons, the resulting field is the same for both types of particles. For linear acceleration 
Eqns. \ref{radfield} and \ref{AfromE} reduce to:
\begin{equation}
\vec{A}_{\parallel}(\theta,t)=\frac{e}{\sqrt{4\pi c}}\left[ \frac{\sin\theta}{(1-\beta\cos\theta)^3}\dot{\beta}_{\parallel}\right]
_{\mathrm{ret}} \hat{e}_{1}
\label{Apar}
\end{equation}
and for perpendicular acceleration:
\begin{equation}
\vec{A}_{\perp}(\theta,t)=\frac{e}{\sqrt{4\pi c}}\left[ \frac{\cos \theta -\beta}{(1-\beta\cos\theta)^3}\dot{\beta}_{\perp}\right]
_{\mathrm{ret}} \hat{e}_{1}
\label{Aperp}
\end{equation}
when $\vec{n}$ lies in the orbital plane of the particle. Note: the subscripts to $\vec{A}$ refer to the type of acceleration
(linear or perpendicular) that cause the vector potential.      
For small $\theta$ these equations can be written in the form:
\begin{equation}
\vec{A}_{\parallel}(\theta,t)=\frac{e}{\sqrt{4\pi c}}\left[ \gamma^{5} \frac{\gamma\theta}{(1+\gamma^{2}\theta^{2})^{3}}
\dot{\beta}_{\parallel}\right]_{\mathrm{ret}} \hat{e}_{1}
\end{equation}
\begin{equation}
\vec{A}_{\perp}(\theta,t)=\frac{e}{\sqrt{4\pi c}}\left[ \frac{1}{2} \gamma^{4} \frac{1}{(1+\gamma^{2}\theta^{2})^{3}}
\dot{\beta}_{\perp}\right]_{\mathrm{ret}} \hat{e}_{1}
\end{equation}
where the relativistic approximation $(1-\beta)^{-1}=2\gamma^2$ is used.
We will now compare the peak values of $A_{\parallel}$ and $A_{\perp}$. The peak of $A_{\parallel}$ lies at $\gamma\theta=1/2$ and the peak
of $A_{\perp}$ at $\theta=0$. We find that:

\begin{equation}
\vec{A}_{\parallel \mathrm{max}} \propto \gamma^{5} \dot{\beta}_{\parallel} \propto \gamma^{2} F_{\parallel}
\end{equation}
\begin{equation}
\vec{A}_{\perp \mathrm{max}} \propto \gamma^{4} \dot{\beta}_{\perp} \propto \gamma^{3} F_{\perp}
\end{equation}
When an electric field is absent the peak value of the vector potential is:
\begin{equation}
\vec{A}_{\mathrm{0,max}} \propto \gamma_{0}^{3} q c B \sin \alpha
\end{equation}
When an electric field $E$ with pitch angle $\delta$ (defined as the angle between $\vec{E}$ and $\vec{\beta}$) is present the peak value of
the vector potential is:
\begin{equation}
\vec{A}_{\mathrm{E,max}} \propto \gamma^{3} q (c B \sin \alpha \pm E\sin\delta) + \gamma^{2} q E \cos\delta
\end{equation}
where the plus sign corresponds to the case where the electric force and the Lorentz force act in the same direction and the minus sign to
the case where
they act in opposite directions. The Lorentz factor $\gamma$ can differ from $\gamma_{0}$ as a result of linear acceleration or deceleration.
Now we can define an amplification factor $N_{\mathrm{amp}}$ as:
\begin{equation}
N_{\mathrm{amp}}=\frac{\vec{A}_{\mathrm{E, max}}}{\vec{A}_{\mathrm{0, max}}}=
\left(\frac{\gamma}{\gamma_{0}}\right)^{3} \left( 1\pm \frac{E\sin\delta}{B c \sin\alpha}\right) + 
\frac{1}{\gamma_{0}}\left(\frac{\gamma}{\gamma_0}\right)^{2} \frac{E\cos\delta}{B c \sin \alpha}
\label{amplification}
\end{equation}
Three effects can be distinguished that cause amplification:
\begin{itemize}
\item The part of the electric field that is directed perpendicular to the particles. The amplification depends on the force as
$N_{\mathrm{amp}}\propto F$.
For a typical case of a horizontal shower in a vertical electric field (see Fig. \ref{trajectories2}) with values
$B=0.5$~G, $\alpha=25^{\circ}$ and $\gamma_{0}=30$ the amplification factor is 1.15 for an electric field of 1~kVm$^{-1}$ and 17 for a
field of 100~kVm$^{-1}$.
\item The part of the electric field that acts linearly on the particles' trajectory. The amplification depends on the force as 
$N_{\mathrm{amp}}\propto F/\gamma_{0}$. For the case of a vertical shower in a vertical electric field and the same characteristic
values as mentioned above the amplification factor is 1.5 for an electric field of 100~kVm$^{-1}$.
\item As the particles are accelerated/decelerated their Lorentz factor increases/decreases. The amplification depends on the Lorentz factor
as $N_{\mathrm{amp}}\propto (\gamma/\gamma_{0})^{3}$. This effect depends strongly on the track lengths of the particles. Suppose a pair is
created in a vertical shower in a vertical electric field and the same characteristic values mentioned above. When the particles have crossed
an altitude difference of 100~m the positron has
reached a Lorentz factor of 50, while the electron's Lorentz factor is down to 10 (see Fig. \ref{trajectories}). The corresponding amplification is $(50^{3}+10^{3})/2\cdot
30^{3} \approx 2.3$. After 200~m the electron has changed direction and its radiation no longer reaches the ground. The positron now has a
Lorentz factor of 70, corresponding to an amplification factor of $6.3$. At a height of 4~km the mean free path length of electrons and
positrons is in the order of a few hundred meter. 
\end{itemize}

\subsubsection*{Amplification of emission from a complete shower} 
So far, only the emission of a single particle pair was discussed. Eqn. \ref{amplification} does not apply to complete showers, consisting of many electron-positron pairs with
varying pitch angles, energies and track lengths. An observer at ground level can see the emission of a particle only for a fraction of its
lifetime, since the particle follows a curved trajectory. When the emission from a particle is amplified due to an electric force, this does
not necessarily mean that an observer will see an increase in emission.
The observed pulses of single particles have a duration of $\Delta t = \pi\rho/c
\gamma^{3}\sim 10^{-11}$~s while the total pulse of a shower is of the order $\Delta T=L/(2 c\gamma^{2})\sim 10^{-8}$~s, where $L$ is the
length of the total shower. The pulses of individual particles are distributed over the period $\Delta T$, so the emission is not totally
coherent. Since the number of particles is much higher than $\Delta T/\Delta t$ the emission is also not completely incoherent.  

The radius of curvature of a relativistic particle is given by Eqn. \ref{radofcurv}. The time width of a single pulse, as measured by a
ground observer, depends on the Lorentz
factor and the applied force as $\Delta t \propto \gamma^{-2} F_{\perp}^{-1}$. Some of the amplification effects mentioned in the section
above will therefore not contribute to a total increase of emission.
\begin{itemize}
\item  The part of the electric force that is directed perpendicular to the particle orbit increases the peak value of $A$ proportional to
$F_{\perp}$. The time width of the observed pulse goes down as $F_{\perp}^{-1}$ and the contribution to the integrated emission of the whole
shower remains the same.
\item The part of the electric force that is directed along the particle orbit does not influence the time width of the pulse (initially). However,
the sign of $A_{\parallel}$ is dependent on the viewing angle (see Eqn. \ref{Aperp}). When the pulses are
distributed over $\Delta T$ pulses of opposite polarity will be added and partially cancel out.
\item When the particle is accelerated its field strength increases proportional to $\gamma^{3}$, while the pulse width decreases as $\gamma^{-2}$.
The contribution to the integrated shower field strength roughly increases with $\gamma$. The power of the integrated pulse increases with
$\gamma^{2}$.
\end{itemize}
From this treatment it appears that the increase in particle velocity is the most important factor in pulse amplification for an integrated
shower. However, when the time width of a pulse decreases because the particles' trajectory is bent over a larger angle this also means
more observers will be able to see emission from that particle. In general, an observer will be able to see a larger fraction of the total 
amount of particles and, in effect, an amplification of the radio pulse.

The total amplification of a radio pulse from a complete shower depends on the distribution of particles and the position of the ground 
observer and cannot be easily predicted. A
detailed Monte Carlo simulation with realistic particle distribution will give more reliable results and is subject of further research. For now, we can regard the amplification
factor as given in Eqn. \ref{amplification} as an upper limit for the observed amplification.

\subsection{Acceleration of ionization electrons} \label{subsec:ionizedelec}
The shower particles ionize air molecules and leave behind free electrons and positive ions. The electrons can recombine with the ions in a time scale of seconds, but on a much
shorter time scale of a few tens of nanoseconds the electrons attach to oxygen molecules forming negative ions \citep{W57}. When an electric field is present
the free electrons are accelerated producing a current pulse. Because of frequent collisions they gain a drift velocity of $\sim 100$ m/s \citep{Ch68}.
The duration of the pulse depends on the attachment time $\tau_\mathrm{att}$ of electrons to oxygen molecules (which is a function
of the electron energy) and the angle under which the shower is viewed. In contrast to the radio emission from the shower electrons/positrons the radiation is not beamed forward
since the free electrons do not become relativistic.
The associated frequency up to which the emission is coherent is $\sim 10$ MHz for an observer in the direction of the shower (inverse of $\tau_\mathrm{att}\sim 100$ ns). 
\begin{figure}[h]
\begin{center}
\centerline{\includegraphics*[width=0.75\columnwidth]{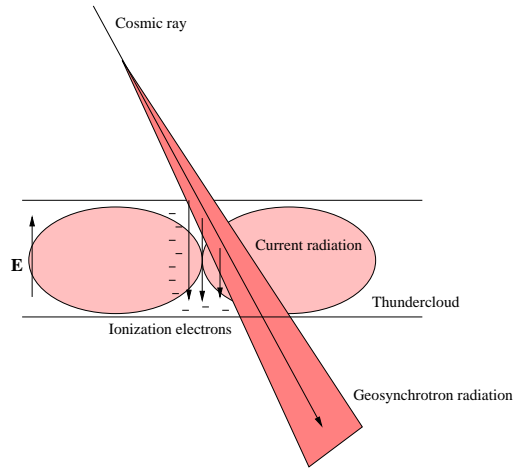}}
\caption{\label{ionization} An EAS passes through a thunderstorm cloud emitting the usual geosynchrotron emission. The ionization electrons are accelerated inside the
cloud. In this picture the radiation pattern is displayed as dipole radiation. The real pattern of a current pulse suffers from diffraction.}
\end{center}
\end{figure}

When ionized electrons gain an energy of $\epsilon>\epsilon_{c}\approx 0.1-1$ MeV they can ionize new molecules. If the electric field is strong enough to accelerate ionization
electrons to such energies a process called runaway breakdown \citep{G92} can occur. The critical field strength of $E_{c}\approx 100-150$ kV/m, needed for this effect, is present only inside
thunderclouds. In the runaway breakdown process two generations of electrons are created: relativistic runaway electrons and slow thermal electrons, which can both contribute to the
radio signal. 
In Fig. \ref{ionization} the contribution of the ionization electrons is schematically displayed. A simple current would produce a dipolar radiation field. The real radiation
pattern will be more complex because of the finite length of the current, the transverse width of the current and the existence of relativistic electrons.
The radiation pattern of the runaway breakdown is calculated in \citet{G02} for a vertical shower and resembles that of a current pulse. The pulse amplitude is calculated to be several orders of magnitude higher
than the geosynchrotron emission from the EAS.

\subsection{Distinguishing between the mechanisms}\label{subsec:disting}
Both mechanisms can be responsible for an amplification of the radio pulse from EAS. There are several ways to distinguish between them:
\begin{enumerate}
\item The radio pulse from the shower electrons will have approximately the same width as produced by the geosynchrotron mechanism and is dependent on the longitudinal separation
of the air shower particles. The pulse width of the ionization current pulse is determined by the electron attachment timescale ($\Delta t_\mathrm{att}\sim 100$ ns) and the timescale
associated with the scale $l_\mathrm{a}$ of the runaway breakdown current $\Delta t \sim l_\mathrm{a}/c \approx 200-300$ ns \citep{G04}. 
\item The pulse from the shower particles will be polarized in the plane of curvature, as in the geosynchrotron mechanism. The polarization
of the ionization current pulse is in the vertical plane of the current and the observer.
\item The radiation from the ionization current is emitted in all directions, while the radiation from the relativistic shower particles is beamed forward, as shown in Fig.
\ref{ionization}.
\end{enumerate}

\section{Experimental Setup}        			   \label{sec:data}

\begin{figure}[h]
\begin{center}
\centerline{\includegraphics*[width=0.8\columnwidth, angle=270]{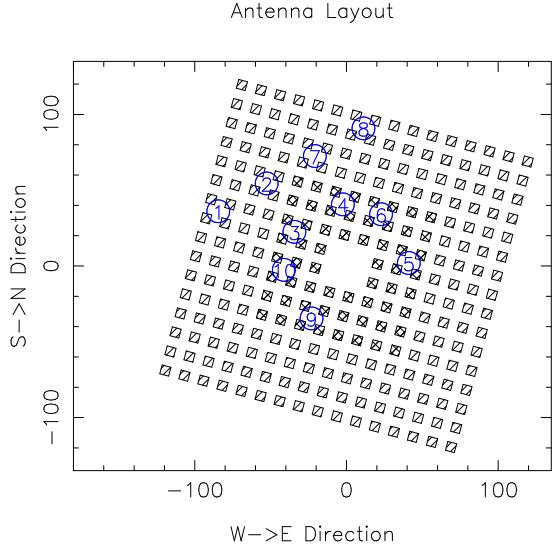}}
\caption{\label{layout} Layout of the LOPES experiment. Boxes indicate positions of KASCADE detector stations, filling up an area of 200~m by 200~m, and circles indicate antenna positions}
\end{center}
\end{figure}

In 2004, the \textsc{lopes} array consisted of 10 dipole antennas, placed on the same location as the \textsc{kascade} experiment \citep{A03} which provides 
triggers for
\textsc{lopes} and records the muon and electron components of the EAS, used to reconstruct the energy of the primary and its direction.
Presently, \textsc{lopes} measures only the polarization in the east-west plane. A layout of the experiment is shown in Fig. \ref{layout}. Details about the experimental setup and the reduction of
the data can be found in
\citet{H04}.  
Additional information about the weather was obtained from a weather station at Karlsruhe (49$^{\circ}$02'N 8$^{\circ}$22'E) in the 
archive of a free weather server\footnote{\texttt{http://meteo.infospace.ru/wcarch/html/index.sht}}. 

Four sets of data were selected from the 2004 database of \textsc{lopes}:
\begin{enumerate}
\item Events with the highest \textsc{kascade} particle count. The events in this selection have either a truncated muon number above $2 \cdot 10^{5}$ or an electron
number exceeding $5 \cdot 10^{6}$ (412 events spread over the period January-September).
\item Fair weather events which took place during periods with 0\% cloud coverage (9455 events spread over the period March-September).
\item Events which took place while the sky was covered by nimbostratus clouds for more than 90\% (2659 events spread over the period January-March).
\item Events during thunderstorms, which were identified by looking at lightning strike maps\footnote{\texttt{http://webcam.paanstra.nl/}} and the dynamic spectra of \textsc{lopes}. The radio emission of actual lightning
strikes show up on these spectra as bright lines because the antenna signal is saturated. In our selection we only used thunderstorms that were visible both on the maps and the
spectra. The radio events that were recorded between these strikes or half an hour before the first or after the last strike on the spectra are regarded as
thunderstorm events (3510 events taken from 11 thunderstorms in the period May-August).
\end{enumerate}

Together, all these events form only a very small fraction of the total \textsc{lopes} database, because the weather information is not complete and even if it was, most weather
conditions do not match the criteria of one of these selections. The selections include events for which a radio signal was not detected. The
weather at the \textsc{lopes} site is expected to differ only slightly from the weather as archived at the Karlsruhe weather station.

To determine whether the radio peak is significant or not a cross-correlation beam was created. 
A combined signal of all \textsc{lopes} antennas is reconstructed by temporal shifting of the pulses in accordance with the arrival direction
in the \textsc{kascade} EAS data. The CC-beam is then calculated by adding correlations of all possible antenna pairs. The radio signal as a function of time is 
fitted with a Gaussian and is considered a detection when the
fitted peak is larger than the background noise by 3 sigma. Since the amplitude calibration of the antennas is not yet completed, the pulse heights are given in arbitrary units.
In the scope of this paper this is not a problem, since we investigate relative differences between sets of events.

\section{Results}       				\label{sec:results}
In \citet{F05} it was shown that the strength of the radio signal depends on the geomagnetic angle as $(1-\cos\alpha)$ when it is 
normalized with the (truncated) muon number. We show this correlation here, using the selection of
events that is the first set listed in Sec. \ref{sec:data} and applying the cuts listed in
Table \ref{tab:cuts} under `largest'. 
\begin{figure}[h]
\begin{center}
\centerline{\includegraphics*[width=0.65\columnwidth, angle=270]{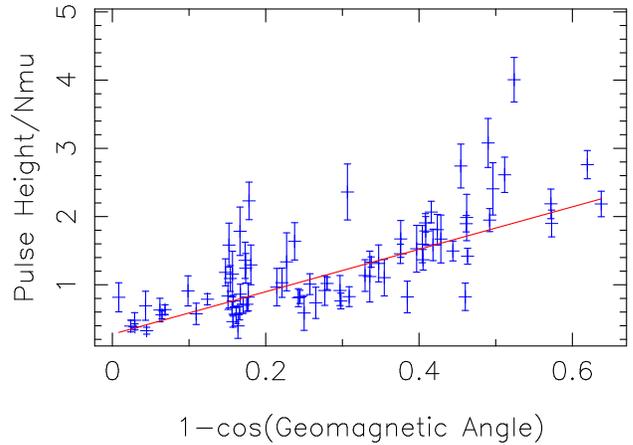}}
\caption{\label{fit} Normalized pulse height of the events from selection 1 of Sec. \ref{sec:data}, plotted against the geomagnetic angle. Excess pulse heights are defined as
the normalized pulse height minus this fit.}
\end{center}
\end{figure}
\begin{table}
\caption{Applied cuts on selections.}
\label{tab:cuts}
\centering
\begin{tabular}{c c c c c}
\hline \hline
cuts             & largest         & conservative    & distant   	& radio   \\
\hline
distance 	 & $<$91\,m        & $<$91\,m        & -                & -		\\
zenith angle 	 & $<50\ ^{\circ}$ & $<50\ ^{\circ}$ & $<50\ ^{\circ}$  & -		\\
no. of antennas  & $\geq 6$        & $\geq 6$        & $\geq 6$         & $\geq 6$  	\\
log(muon no)     & $>5.3$          & $>5$            & $>5$ 	        & -		\\
radio peak       & $>3\sigma$	   & $>3\sigma$      & $>3\sigma$       & largest  \\
\hline
\end{tabular}
\end{table}
In Fig. \ref{fit} the normalized pulse heights are plotted against geomagnetic angle and a fit is made. The normalization is done by dividing
by the truncated muon number and multiplying by $10^{6}$. In order to compare data points to this fit we will calculate the difference between the
points and the fit, not the ratio. The reason for this is that the ratio is a value that is normalized for geomagnetic value. For an
amplified pulse caused by an electric field this normalization is not suitable. The total radiation can be seen as consisting of a geomagnetic
part (which scales with geomagnetic angle) and an electric field part (which does not scale with geomagnetic angle). 
We therefore define the pulse 
height excess as the \emph{normalized} pulse height minus the fit value (the difference between a data point and the fit in
Fig. \ref{fit}). The
error-weighed mean pulse height excess is calculated as:
\begin{equation}
\chi=\frac{1}{N}\displaystyle\sum^{N}_{i=1}\frac{Y_{i}}{\sigma_{i}}=0.33
\end{equation}
where $N=\sum\sigma_i^{-1}$, $Y_{i}$ are the pulse height excess values and $\sigma_{i}$ the errors. Various effects are included 
in the calculation of this error:
\begin{equation}
\sigma_i^2=\sigma_{bg}^2+\sigma_{fit}^2+\sigma_{phase}^2+\sigma_{ge}^2
\end{equation}
where $\sigma_{bg}$ is the root mean square of the background signal and $\sigma_{fit}$ is the error in the Gaussian fit to the
measured radio pulse. An error in the phase calibration of the antennas translates into an error $\sigma_{phase}$ in the formed CC-beam.   
The antenna gain factor depends on the signal direction, so an error in the signal direction translates into an additional error in the gain factor,
$\sigma_{ge}$. The last error is very small for zenith angles below $50^{\circ}$. For most radio pulses $\sigma_{bg}$ is dominant, but for strong pulses
$\sigma_{phase}$ gives the
largest contribution. All but one data point in Fig. \ref{fit} have an excess $<2$.
\begin{figure}[h]
\begin{center}
\centerline{\includegraphics*[width=0.65\columnwidth, angle=270]{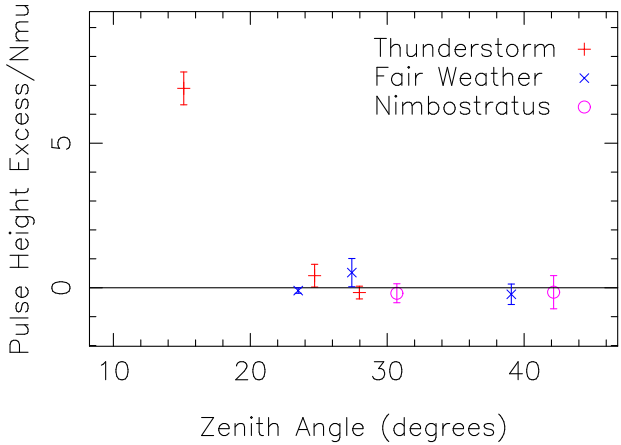}}
\caption{\label{conservative} Pulse height excess is plotted against zenith angle for conservative cut.}
\end{center}
\end{figure}
In the evaluation of the fair weather, nimbostratus and thunderstorm events, three cuts were applied, which are all listed in Table \ref{tab:cuts}. In the `conservative' cut 
the maximum distance of the shower core to the array centre is 91 m and the zenith angle of the shower must be smaller than 50$^{\circ}$. Beyond these limits the 
values of the general shower parameters may not be well reconstructed by \textsc{kascade} (see \citet{A03}). Furthermore, at least 6 out of 10 antennas must have detected a radio signal and the reconstructed muon number must exceed
$10^5$. In the `distant' cut the constraint on distance is dropped. 
The conservative cut yields only 3 thunderstorms events, 3 fair weather events and 2 nimbostratus events. Their pulse height excesses are plotted against zenith angle in Fig.
\ref{conservative}. One of the three thunderstorm events shows a large excess, while the other events deviate from the fit within 1$\sigma$. Because of the low statistics we will
focus on the results of the `distant' cut, which leaves 14 out of 3510 thunderstorm events (0.40\%), 15 out of 9455 (0.16\%) fair weather events and 7 out of 2659 (0.26\%)
nimbostratus events. The detection rate increases considerably during thunderstorm. The detection rate during nimbostratus conditions also seems to be slightly higher, but
one should note that this is all low number statistics.
\begin{table}
\caption{Mean pulse height excess.}
\label{tab:spread}
\centering
\begin{tabular}{c c c c}
\hline \hline
&cloudless&nimbostratus&thunderstorm\\
\hline
$\chi$&0.17&-0.15&1.27\\
\hline
\end{tabular}
\end{table}

\begin{figure}[h]
\begin{center}
\centerline{\includegraphics*[width=0.65\columnwidth, angle=270]{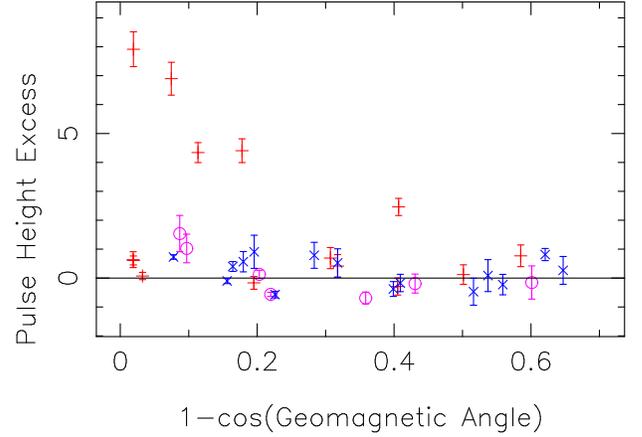}}
\caption{\label{exc-vs-angB} Pulse height excess is plotted against geomagnetic angle for distant cut.}
\end{center}
\end{figure}

\begin{figure}[h]
\begin{center}
\centerline{\includegraphics*[width=0.65\columnwidth, angle=270]{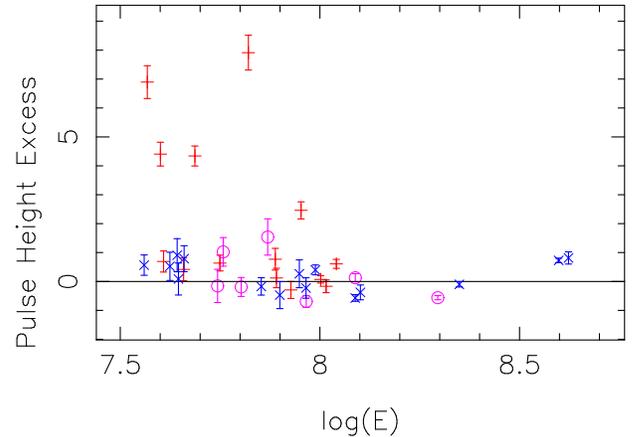}}
\caption{\label{exc-vs-logE} Pulse height excess is plotted against shower energy for distant cut.}
\end{center}
\end{figure}

\begin{figure}[h]
\begin{center}
\centerline{\includegraphics*[width=0.65\columnwidth, angle=270]{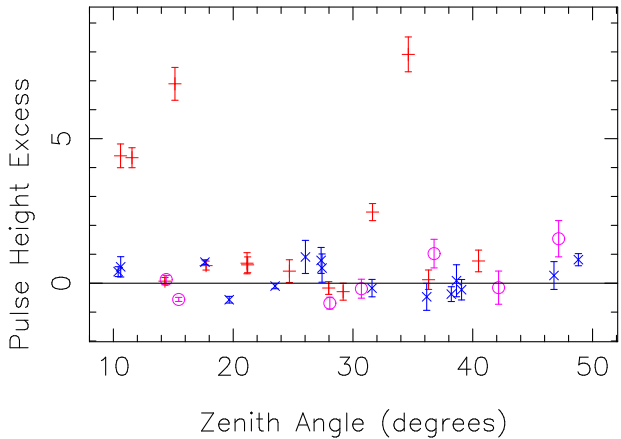}}
\caption{\label{exc-vs-zen} Pulse height excess is plotted against zenith angle for distant cut.}
\end{center}
\end{figure}

\begin{figure}[h]
\begin{center}
\centerline{\includegraphics*[width=0.65\columnwidth, angle=270]{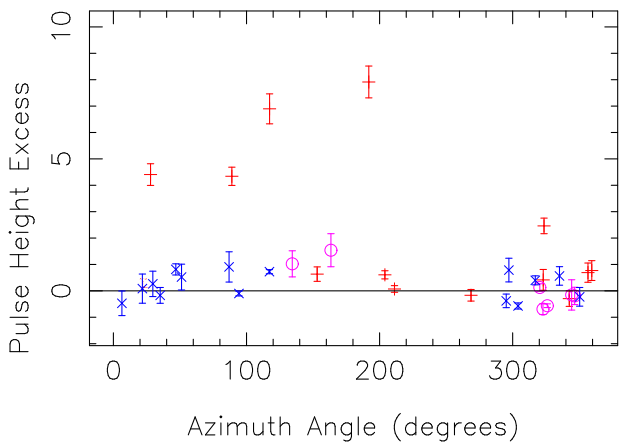}}
\caption{\label{exc-vs-az} Pulse height excess is plotted against azimuth angle to the shower axis for distant cut.}
\end{center}
\end{figure}

\begin{figure}[h]
\begin{center}
\centerline{\includegraphics*[width=0.65\columnwidth, angle=270]{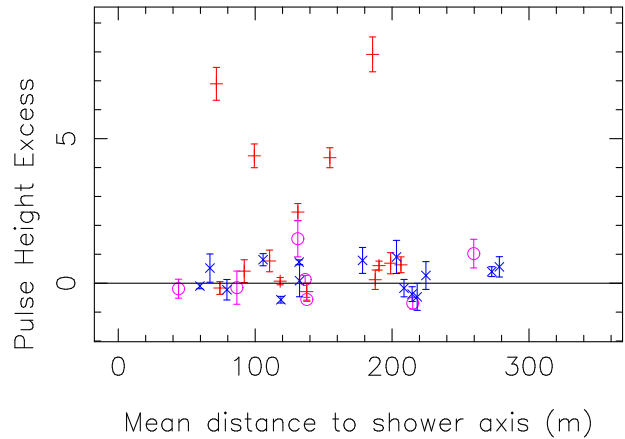}}
\caption{\label{exc-vs-dist} Pulse height excess is plotted against mean distance to the shower axis for distant cut. \textsc{kascade} reconstruction of muon number becomes
unreliable above 91~m.}
\end{center}
\end{figure}

In Fig. \ref{exc-vs-angB} the excess pulse height for events of different weather selections are plotted against geomagnetic angle, where the distant cut is applied. 
Figs. \ref{exc-vs-logE}, \ref{exc-vs-zen}, \ref{exc-vs-az} and \ref{exc-vs-dist} contain the same data points, now plotted against 
respectively EAS energy as estimated by \textsc{kascade}, zenith 
angle, azimuth angle and mean distance of the antennas to the shower axis. There is a bias towards events with a positive pulse height excess, because these have a higher chance to be
detected on a 3$\sigma$ level.
The mean pulse height excesses are listed in Table \ref{tab:spread}.


\begin{figure}[h]
\begin{center}
\centerline{\includegraphics*[width=0.65\columnwidth, angle=270]{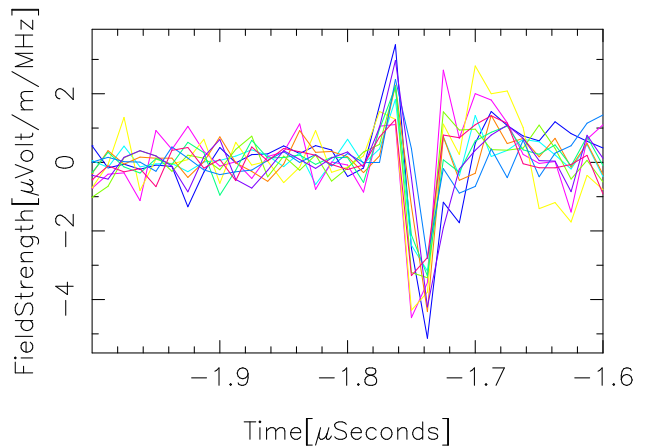}}
\caption{\label{guiAll} Signals of ten \textsc{lopes} antennas for a very radio bright event (event K in Fig. \ref{large-vs-zen}). (Field
strength values are not calibrated.)}
\end{center}
\end{figure}\begin{figure}[h]
\begin{center}
\centerline{\includegraphics*[width=0.65\columnwidth, angle=270]{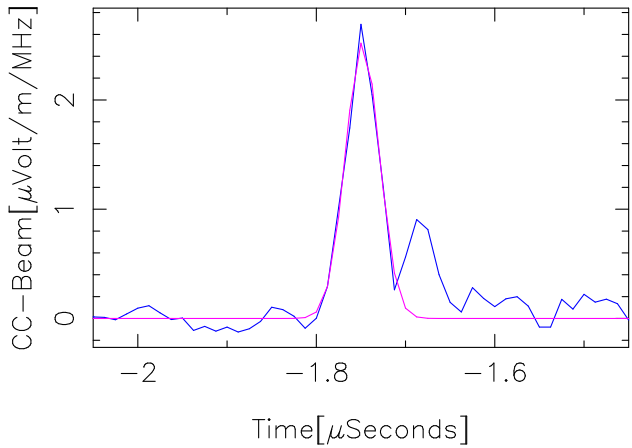}}
\caption{\label{guiCC} Cross correlation beam of the ten signals of Fig. \ref{guiAll}. The blue line is the cross correlated signal and the 
purple line is a Gaussian fit. (Field
strength values are not calibrated.)}
\end{center}
\end{figure}

\begin{figure}[h]
\begin{center}
\centerline{\includegraphics*[width=0.65\columnwidth, angle=270]{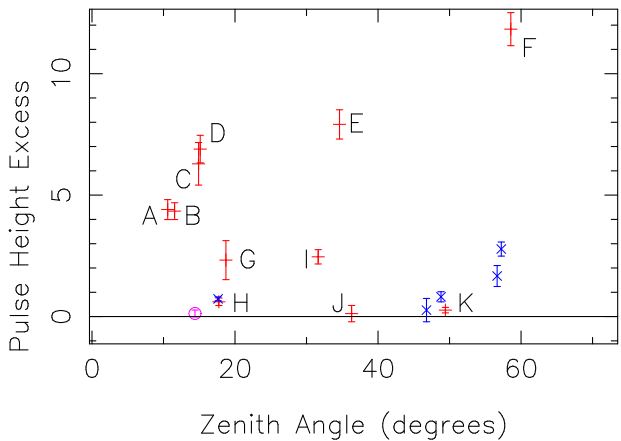}}
\caption{\label{large-vs-zen} Pulse height excess is plotted against the zenith angle for radio cut. The thunderstorm events are labelled A through K.}
\end{center}
\end{figure}

Some events have extremely nice radio pulses like in Fig. \ref{guiAll}. The signal-to-noise ratio in the cross-correlation beam (see Fig. \ref{guiCC}) is much larger than for other
events and the coherence is very high. A selection of these events was made by eye (`radio' cut in Table \ref{tab:cuts}) and the pulse height excesses are plotted in Fig.
\ref{large-vs-zen}. All of these events have signal-to-noise ratios of $>10$. In the fair weather selection, 5 out of 9455 events had such strong radio emission (0.05\%), in the nimbostratus selection 1 out of 2659 (0.04\%) and in
the thunderstorm selection 11 out of 3510 events (0.3\%). 
The events that show a bright pulse because of the shower size appear near the bottom of the plot, because of the normalization with muon number. 
Bright pulses from showers with a relatively small muon number appear in the upper part of the plot. Only thunderstorm events are present in this
region.

\begin{figure}[h]
\begin{center}
\centerline{\includegraphics*[width=0.65\columnwidth, angle=270]{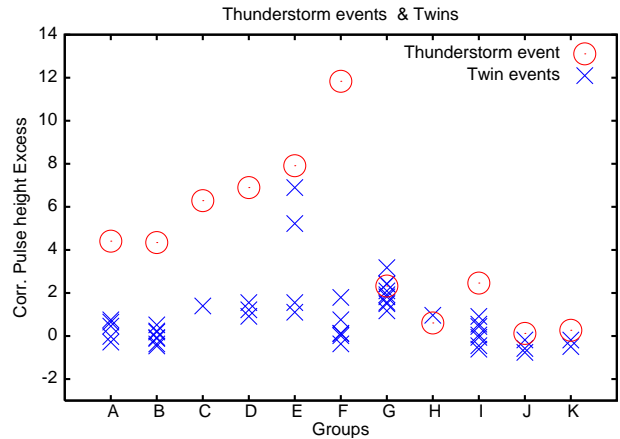}}
\caption{\label{twins} Pulse height excess, corrected for geomagnetic angle, for thunderstorm events A through K and their twins. In group E, two of the twin events are probably also thunderstorm events (see text for details).}
\end{center}
\end{figure}

To further check the uniqueness of the high excess thunderstorm events, new selections of twin events were made from the \textsc{lopes} database. For each thunderstorm event (A
through K in Fig. \ref{large-vs-zen}) a
selection was made of events with approximately the same muon and electron number (both within 5\%). The zenith angle and mean distance to the shower axis of these twin events is
not necessarily the same. The pulse height excesses, corrected for geomagnetic angle by dividing by the fit in Fig. \ref{fit}, of all events are plotted in Fig. \ref{twins} by
group. For groups A through F the pulse height excess of the thunderstorm event is significantly larger than those of their twins, while in groups G through K the thunderstorm events have
excesses similar to those of their twins. The pulse heights are not normalized for mean distance to the shower core, since this dependence is not yet clearly established. Any reasonable
normalization (e.g. $\propto \exp(-R/100$ m)) will not change the appearance of Fig. \ref{twins} significantly, i.e. the thunderstorm events of groups A through F still have much
larger pulse height excesses than their twins.

All the events that show a large excess in Fig. \ref{large-vs-zen} have significantly larger pulse heights than their twins, while the
low excess events have pulse heights similar to their twins. In group E two of the twin events have a large pulse height excess. Lightning maps and data from the weather station
show that these events have also occured under thunderstorm conditions. (They were not in the original selection because no lightning strikes were visible in the dynamic spectrum
during these thunderstorms.)
\begin{figure}[h]
\begin{center}
\centerline{\includegraphics*[width=0.65\columnwidth, angle=270]{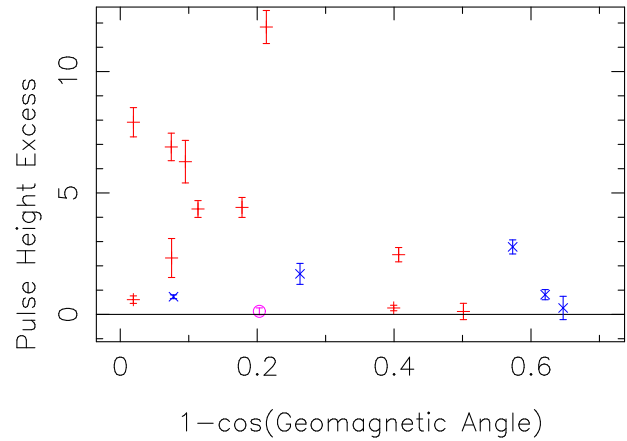}}
\caption{\label{large-vs-angB} Pulse height excess is plotted against the geomagnetic angle for radio cut.}
\end{center}
\end{figure}

\section{Discussion}                \label{sec:discussion}
It is found that during thunderstorms the radio signal from EAS can be strongly amplified. Due to the low number of events for which the
weather conditions could be reliably reconstructed, it was necessary to include events that were more than 91~m away from the array
core.
The \textsc{kascade} reconstruction of the muon number is not fully reliable for showers with large zenith angle or a distant core. When 
the estimation of the muon number is too
low the calculated excess values will be too high. The conservative cut is the most reliable but leaves the fewest data points. The
distant cut leaves more data points simply because there are many events with a distant core, not because they are detected more
efficiently. Although 
the number of
data points in the conservative cut is too low for statistics, it reflects the structure of the plots where the distant cut is applied. Large excesses for thunderstorm events were
found, while the spread
for fair weather and nimbostratus events is small. 

Although some thunderstorm events show a large excess, others fall inside the same spread as fair weather and nimbostratus events. This can be because the selected time windows
for the thunderstorm selection probably contains some time before and after the thunderstorms. Also, even when the thunderstorm is at its strongest, the amplification of the
radio emission will depend on the local field distribution inside the cloud and the angle of the shower axis with the electric field.  

During nimbostratus conditions no amplified radio emission was found. This could be because the electric field is too weak, or because these clouds have a 
smaller vertical extent than
thunderstorm clouds and most showers reach their maximum above the nimbostratus cloud, where no large electric field is present.
 
Figs. \ref{exc-vs-logE} through \ref{exc-vs-dist} show no correlations between the pulse height excess and other EAS parameters (primary energy, zenith angle, azimuth angle and distance to the shower axis), which indicates that the observed amplification is caused by
the weather condition. Fig. \ref{large-vs-angB} and perhaps Fig. \ref{exc-vs-angB} seem to indicate that the amplified pulses cluster at low geomagnetic angles.
The statistics are not good enough to make a statement on whether this feature is coincidental or not. More data is needed to study this possible correlation.

The twin events in Fig. \ref{twins} have the same muon and electron number, but other EAS parameters vary. The values
in this plot are normalized with muon number and corrected for geomagnetic angle. The spread in pulse height excess within a group of twin
events can be due to differences in zenith angle, azimuth angle or distance. The spread is, in groups A through F, small compared to the pulse height
excess value of the thunderstorm event(s). When, in the future, a larger database of events is available a similar analysis can be done for twin
events that also share the same zenith angle, azimuth angle and distance.

Radio pulses observed with \textsc{lopes} typically have a width of $50-60$ ns. Due to the 40--80 MHz band filtering any radio pulse shorter than that will appear broadened. 
Broader radio pulses, however, do maintain more or less their original width. 
The observed pulses of the amplified thunderstorm events have widths of $\sim 50$ ns (see e.g. Fig. \ref{guiCC}) like all other \textsc{lopes} events, and are
probably not ionization current pulses, which would have widths of at least 100 ns (see Sec. \ref{subsec:disting})
We suspect therefore that it is the direct influence of the electric field on the shower electrons and positrons that is responsible for the amplification of the radio emission
during thunderstorm conditions. 

The small distances between the antennas does not allow an evaluation of the lateral distribution of the radio emission and \textsc{lopes}
only measures one polarization, so no additional tests can be done to identify the mechanism at the moment.

\section{Conclusion}                \label{sec:conclusion}
It is shown that during thunderstorm conditions the radio emission of EAS is largely amplified. There are two mechanisms which can explain this amplification:
acceleration radiation from the shower electrons and radiation from a current pulse of (runaway) ionization electrons. 
The measured pulse widths ($\sim 50$~ns) suggest that the latter cannot be the observed mechanism.
To identify the mechanism with more certainty, more information about the
real pulse width, lateral distribution and polarization is needed. At the moment, the \textsc{lopes} experiment is unable to provide this information, but with future
additions to the experiment, such as dual polarization, it will be possible to resolve this problem. Also, in the short future,
\textsc{lofar} stations will be able to help find the answer, since they occupy a larger ground area, operate in a wider frequency range and measure polarization. 

For both the \textsc{lopes} and the \textsc{lofar} experiment it is advisable to keep detailed weather information, like cloud coverage, ground level electric field and the
occurence of lightning strikes. 

With \textsc{lofar} it will be possible to trace lightning activity by three dimensional imaging. This technique allows localization and 
mapping of lightning strikes, 
and possibly also thunderstorm processes emitting weaker radiation such as stepped leaders and high altitude lightning. This offers unique 
opportunities and promises significant further advances in this area.
 
\begin{acknowledgements}
LOPES was supported by the German Federal Ministry of Education and Research. 
The KASCADE-Grande experiment is supported by the German Federal Ministry of 
Education and Research, The MIUR of Italy, te Polish Ministry of Science and 
Higher Education and the Romanian Ministry of Education and Research (grant 
CEEX~05-D11-79/2005).
\end{acknowledgements}

\bibliographystyle{aa}
\bibliography{6006}

\end{document}